# Innovative design and construction technique for the Cilindrical GEM detector for the BESIII experiment

A. Amoroso[5,10], M. Alexeev[5,10], R. Baldini Ferroli[1,3], M. Bertani[3], D. Bettoni[2], F. Bianchi[5,10], A. Calcaterra[3], N. Canale[3], M. Capodiferro[3], V. Carassiti[2], S. Cerioni[3], JY. Chai[1,7], S. Chiozzi, G. Cibinetto[2], F. Cossio[5,10], A. Cotta Ramusino[3], F. De Mori[5,10], M. Destefanis[5,10], J. Dong[1], F. Evangelisti[2], R. Farinelli[2,8], L. Fava[5,11], G. Felici[3], E. Fioravanti[2], I. Garzia[2,8], M. Gatta[3], M. Greco[5,10], CY. Leng[5,7], L. Lavezzi[1,5], H. Li[5,10], M. Maggiora[5,10], R. Malaguti[2], S. Marcello[5,10], M. Melchiorri[2], G. Mezzadri[2,8], M. Mignone[5], G. Morello[3], S. Pacetti[9], P. Patteri[3], J. Pellegrino[5,10], A. Pelosi,[3] A. Rivetti[5], M. D. Rolo[5], M. Savriè[2,8], M. Scodeggio[2,8], E. Soldani[3], S. Sosio[5,10], S. Spataro[5,10], E. Tskhadadze[3], S. Verma[2], R. Wheadon[5], L. Yan[5,10]

[1]Institute of High Energy Physics, Chinese Academy of Sciences, Beijing, China; [2]INFN, Sezione di Ferrara, Italy; [3]INFN, Laboratori Nazionali di Frascati, Frascati (Roma), Italy; [4]INFN, Sezione di Perugia, Perugia, Italy, INFN, Sezione di Roma, c/o Università La Sapienza, Roma, Italy; [5]INFN, Sezione di Torino, Torino, Italy; [6]Joint Institute for Nuclear Research (JINR), Dubna, Russia; [7]Politecnico di Torino, Dipartimento di Elettronica e Telecomunicazioni, Torino, Italy; [8]Università di Ferrara, Dipartimento di Fisica, Ferrara, Italy; [9]Università di Perugia, Dipartimento di Fisica e Geologia, Perugia, Italy; [10]Università di Torino, Dipartimento di Fisica, Torino, Italy; Università Piemonte Orientale, Alessandria, Italy

**Abstract.** Gas detector are very light instrument used in high energy physics to measure the particle properties: position and momentum.
Through high electric field is possible to use the Gas Electron Multiplier (GEM) technology to detect the particles and to exploit the its properties to construct a large area detector, such as the new IT for BESIII. The state of the art in the GEM production allow to create very large area GEM foils (up to 50x100 cm2) and thanks to the small thickness of these foil is it possible to shape it to the desired form: a Cylindrical Gas Electron Multiplier (CGEM) is then proposed.

The innovative construction technique based on Rohacell, a PMI foam, will give solidity to cathode and anode with a very low impact on material budget. The entire detector is sustained by permaglass rings glued at the edges. These rings are use to assembly the CGEM together with a dedicated Vertical Insertion System and moreover there is placed the On-Detector electronic. The anode has been improved w.r.t. the state of the art through a jagged readout that minimize the inter-strip capacitance.

The mechanical challenge of this detector requires a precision of the entire geometry within few hundreds of microns in the whole area.
In this presentation will be presented an overview of the construction technique and the validation of this technique through the realization of a CGEM and its first tests.
These activities are performed within the framework of the BESIIICGEM Project (645664), funded by the European Commission in the action H2020-RISE-MSCA-2014.

## 1 CGEM Project

The existing Inner Tracker detector of the Beijing Spectrometer BESIII [1], the Multi-layer Drift Chamber (MDC), is showing aging effects due to the high luminosity reached by the Beijing Electron Positron Collider II (BEPCII).

The proposed solution of a new IT consists in a cylindrical triple-GEM detector (CGEM) composed by five concentric electrodes: the cathode, three GEM foils, and the anode (Fig. 1). The spacing between the cathode and the first GEM foil (conversion gap) is 5 mm, while all the other gaps are 2 mm.

Each GEM foil is made by a thin Kapton foil of 50 nm, copper clad on each side, with a high surface density of holes [2]. The holes have a bi-conical structure with external (internal) diameter of 70 nm (50 nm). A voltage of 270V is applied between the two copper sides in order to produce a strong field in the holes, of the order of 54 kV/cm, which multiplies the number of electrons produced by a charged particle crossing the detector. The triple-GEM configuration allows to reach high gains minimizing the discharge probability.

Due to the limited space available in the inner part of the BESIII spectrometer, the mechanical design of the detector must be very compact and properly suited, including the on-board electronics.

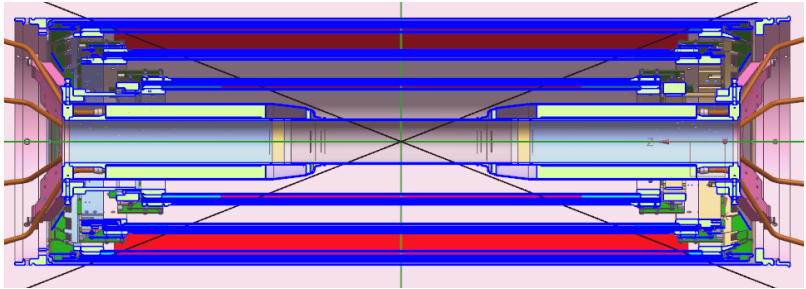

**Fig. 1.** CGEM detector into the BES3 spectrometer. The three CGEM layer are blue

## 2 New Development Feature

To ensure the low material budget requested for the BES3 inner tracker the Anode and Cathode supports are made with ROHACELL, a low density foam that ensures, temperature resistance up to 220°C. The closed cell structure, make ROHACELL foam ideal for high performance sandwich structures combining low multiple scattering and rigidity properties.

The cylindric layout is given by permaglass rings glued on the edge of the electrodes. They provide the designed gap between the difference GEM foils. The rings are shaped as well for the gas system and as support structure for a complete chamber.

The CGEM detectors are designed to provide two spatial coordinates for each chamber. The Anode layout collects data for $x$ and $v$ coordinates (skewed of about 46°, -31°, and 33° [4], for the three chambers respectively), allowing to disentangle multiple hits scenarios. In order to reduce the inter-strip capacitance a Jagged Strip Anode, with *x* strip pitch shrinked in coincidence of *v* strip crossings reduce the inter-strip capacitance of about 30%.

Due to the expected high signal rates the ASIC technology as to be exploited. A new ASIC chip named TIGER (Torino Integrated GEM Electronics for Readout) is developed for the electronic readout.

The new features of CGEM are shown in Fig. 2.

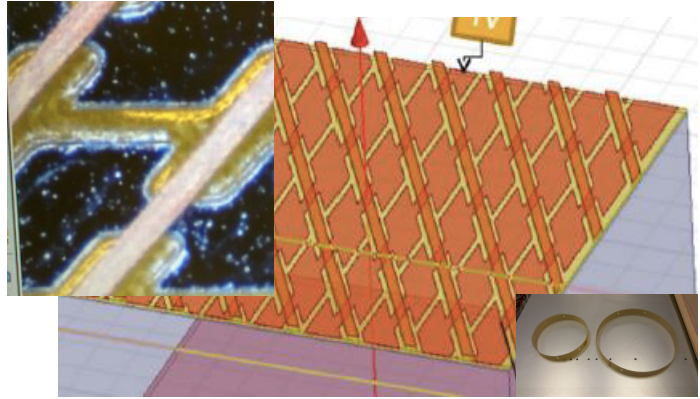

**Fig. 2.** New developed features from CGEM project. Right: permaglass rings. Center: schematic view of the coordinates setup. Left: Jagged strip view.

## 3    Building method

The KLOE-2 experiment has developed a special assembling technique, in order to obtain cylindrical GEM electrodes [3]. The maximun dimension of a Kapton foil is 60cm: this dimension is not sufficient to cover the CGEM layer 2 and 3 required surface. In this scenario, two GEM foils are glued together on a plane to obtain a single larger foil. We decided to use Araldite-103, a multipurpose, two component, room temperature curing, transparent liquid adhesive of high strength, to glue the GEM foils.

A small quantity of glue (1.5mm strip) on the overlap area (2.8mm) gives safety margin to prevent that glue pour out from the overlap region on the sensitive area. In order to ensure a uniform pression on the glued region, a vacuum bag is applied. To reach the glue fully curing 24 hours are needed.

The 3 GEM foils are cylindrically shaped on aluminum mandrel coated with a 400 nm thick Teflon film, which provides a non-sticky, low friction surface. The two Permeglass rings are glued on the edges of the electrode and the foils are glued on their opposite side. Finally, the mandrel is inserted on a vacuum bag until the complete curing of the glue.

The cathode and anode are constructed with a similar procedure, but using the ROHACELL as support structure.

The final assembly is done using a Vertical Insertion Machine (Clessidra) and the assembly of the five electrodes proceeds from the outermost electrode (the anode) to the innermost one (the cathode). Thanks to the Teflon surface, the cylindrical electrode can be easily extracted from the mandrel. The extraction procedure operated with the "Clessidra" is shown in Fig. 4.

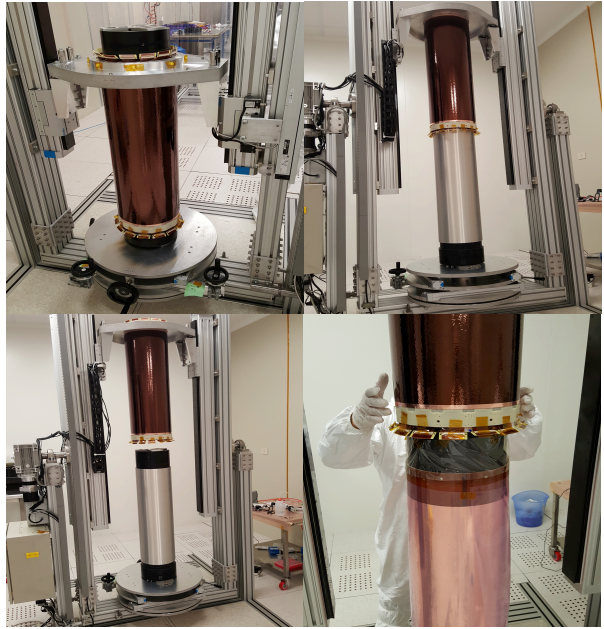

**Fig. 3.** Assembly procedure with the Vertical Insertion Machine

## 4     Conclusions

A new CGEM construction procedure was discussed. The method implies the use of ROAHCELL foam, permaglass rings, and the dedicated onboard electronics. The full-size layer 2 prototype and the layer 1 were already constructed and the latter is now at test beam stage at CERN H4 beam test facility.